\documentclass[a4,12pt,epsf]{article}

\usepackage{psfig}

\begin{document}
%%%%% title %%%%%
\begin{center}{\Large \bf
Electroweak Penguin and Leptophobic $Z^\prime$ model
}
\end{center}

%%%%% author(s) %%%%%
\begin{center}
S. Baek$^a$
J. H. Jeon$^a$
C. S. Kim$^a$\footnote{Talk given at Summer Institute 2006, APCTP Pohang Korea, August 23-30, 2006}
and
Chaehyun Yu$^b$\footnote{chyu@korea.ac.kr}
\vspace{6pt}\\

%%%%% address(es) %%%%%
$^a$
{\it
Department of Physics, Yonsei University,
Seoul 120-479, Korea
}
% Anoterh Address
%{
\\
$^b$
{\it
Department of Physics, Korea University,
Seoul 136-701, Korea
}
\end{center}
%%%%% abstruct %%%%%
\begin{abstract}
We consider the leptophobic $Z^\prime$ model which can appear naturally
in the flipped SU(5) or string-inspired $E_6$ models.
This model can be constrained by measurements of
 the $B\to M \nu\bar{\nu}$ decays and $\Delta m_s$.
We find that although the latter give  much stronger constraints
on the coupling than the former, they are complementary to each
other.
\end{abstract}

%%%%%%%%%%%%%%%%%%%%
\section{Introduction}

Since in the standard model (SM)
the flavor changing neutral current (FCNC) processes
appear at the quantum level
with suppression factors by small electroweak gauge coupling,
CKM matrix elements, and loop momenta,
they are very sensitive to probe new physics (NP) beyond the SM
which have an enhancement factor in the coupling
or have tree-level FCNCs.

The decay of $B$ mesons accumulated largely at asymmetric $B$-factories
and Tevatron give an opportunity to probe NP models via the rare $B$ decays
induced by FCNCs.
Recently, among several sources for FCNCs in the $B$ decays,
the electroweak (EW) penguin operators have drawn much interest.
For example, the QCD penguin dominant $B\to K\pi$ decays
appear to be very interesting
since branching ratios (BRs) and mixing-induced CP asymmetry
allow much room for large NP contribution, especially
in the EW penguin sector~\cite{Kim:2005jp,Wu:2006ur}.

Most of models contributing to
the EW penguin sector have a severe constraint from
the $b\to s\gamma$ decay.
While, models such as the $Z^\prime$ model are free from such constraints
although they predict the EW penguin contributions.
In order to probe such NP models, one must resort to nonleptonic decays
or very rare process $B\to M \nu\bar{\nu} (M=\pi,K,\rho,K^\ast)$.
However, nonleptonic decays might be inefficient since they suffer from
large hadronic uncertainties and EW penguins contributions are subdominant
in nonleptonic decays.

Recently, D{\O}~\cite{D0} and CDF~\cite{Gomez-Ceballos:2006qm} Collaborations at
Fermilab Tevatron have reported
the first observation of the mass difference $\Delta m_s$
in the $B_s^0 -\overline{B}_s^0$ system which induced by the $b\to s$ FCNC:
 \begin{eqnarray}
  \textrm{D{\O}}~&:&~\quad 17 ~ \textrm{ps}^{-1} < \Delta m_s < ~21 ~ \textrm{ps}^{-1}
                        ~~\left( 90 \% ~ \textrm{C.L.}\right) ,
  \nonumber\\
  \textrm{CDF}  ~&:&~\quad \Delta m_s = 17.33_{-0.21}^{+0.42} (\textrm{stat.})
                                    \pm 0.07 (\textrm{syst.})~ \textrm{ps}^{-1}.
 \end{eqnarray}
These measurements may give strong constraints on the NP models,
which predict $b \to s $ FCNC transitions~\cite{MSSM,RS-kim}.

In the present work, we focus on the leptophobic
$Z^\prime$ model motivated from the flipped SU(5) or string-inspired
$E_6$ models as a viable NP model. In Sec.~\ref{sec2},
we briefly introduce the leptophobic $Z^\prime$ model.
Section~\ref{sec3} deals with
$B \rightarrow M \nu \bar{\nu}~ (M= \pi, K, \rho, K^*)$ decays
within the leptophobic $Z^\prime$ model.
We investigate implications of $\Delta m_s$ measurements on this model
in Sec.~\ref{sec4} and conclude in Sec.~\ref{sec5}.

\section{Leptophobic $Z^\prime$ model and FCNC
\label{sec2}}

In many new physics scenarios containing an additional $U(1)^\prime$ gauge
 group at the low energy, the new neutral gauge boson $Z^\prime$
would have a property of leptophobia, which means that
the $Z^\prime$ boson does not couple to the ordinary SM charged leptons.
In flipped SU(5)$\times$U(1) scenario~\cite{Lopez:1996ta},
leptophobia of the $Z^\prime$ boson can be given naturally because
the neutrino is subject to the different representation with the charged
leptons.
 Another scenario for leptophobia can be found in the $E_6$ model with
kinetic mixing, where in this model leptophobia is somewhat accidental.
After breaking of the $E_6$ group,
the low energy effective theory contains an extra $\textrm{U(1)}^\prime$
which is a linear combination of $\textrm{U(1)}_\psi$ and $\textrm{U(1)}_\chi$
with a $E_6$ mixing angle $\theta$ \cite{Rizzo:1998ut}.
Then, the general interaction Lagrangian of fermion fields and $Z^\prime$
gauge boson can be written as
\begin{equation}
{\cal L}_{\rm int} = - \lambda \frac{g_2}{\cos \theta_W}
\sqrt{\frac{5 \sin^2 \theta_W}{3}}
\bar{\psi} \gamma^\mu \left( Q^\prime + \sqrt{\frac{3}{5}}\delta Y_{SM} \right)
\psi Z_\mu^\prime ~,
\end{equation}
where the ratio of gauge couplings $\lambda = g_{Q^\prime}/g_Y$,
and $\delta=-\tan \chi/\lambda$ \cite{Rizzo:1998ut}.
Since the general fermion-$Z^\prime$ couplings depend on two free parameters,
$\tan \theta$ and $\delta$, effectively,
the $Z^\prime$ boson can be leptophobic within an appropriate embedding of
the SM particles \cite{Rizzo:1998ut,Leroux:2001fx}.

Assuming  $V_L^d = 1$ in the $E_6$ model and  flipped SU(5) model,
only $Z^\prime$-mediating FCNCs in the right-handed down-type quarks
survive.
Then, one can get the FCNC Lagrangian for the $b\to q (q=s,d)$
transition~\cite{Jeon:2006nq}
\begin{equation}
{\cal L}_{\rm FCNC}^{Z^\prime} = - \frac{g_2}{2 \cos \theta_W}
U_{qb}^{Z^\prime} \bar{q}_R \gamma^\mu b_R Z_\mu^\prime ,
\end{equation}
where all the theoretical uncertainties including the mixing
parameters are absorbed into the coupling $U_{qb}^{Z^\prime}$. The
coupling $U_{sb}^{Z^\prime}$ has in general CP violating complex
phase, which we denote as $\phi_{sb}^{Z^\prime}$. We note that the
leptophobic $Z^\prime$ boson is not well constrained by
experiments including the charged leptons such as $b\to s \ell^+
\ell^-$ or $B_{(s)} \to \ell^+ \ell^-$, while the typical new
physics models are strongly constrained by such experiments.

\section{Exclusive $B\to M\nu\bar{\nu} $ Decays
\label{sec3}}

In this section, we consider the $B\to M\nu\bar{\nu} $
decays in the leptophobic $Z^\prime$ model.
The $B\to M\nu\bar{\nu}$ decays are measured via the scalar or vector meson with
the missing energy signal.

Theoretical estimates for BRs of
the $B\to M\nu\bar{\nu} $ decays
in the SM  are $0.22^{+0.27}_{-0.17}$, $5.31^{+1.11}_{-1.03}$,
$0.49^{+0.61}_{-0.38}$, and $11.15^{+3.05}_{-2.70}$ in units of $10^{-6}$,
respectively.
While experiments by the Belle and BaBar Collaborations have reported
only upper limits on BRs of
$B\to K\nu \bar{\nu}$ and $B\to \pi \nu \bar{\nu}$ decays
\cite{Abe:2005bq,Aubert:2004ws},
where the experimental bounds are about 7 times larger than the SM expectation
for the $K$ production and much larger by an order of $10^3$ for the $\pi$
production.

%%%%%%%%%%%%%%%%%%%%%%%%%%%%%%%%%%%%%%%%%%  Fig. 1
%%%%%%%%%%%%%%%%%%%%%%%%%%%%%%%%%%%%%%%%%%
\begin{figure}
\begin{center}
\begin{tabular}{cc}
~~~\psfig{file=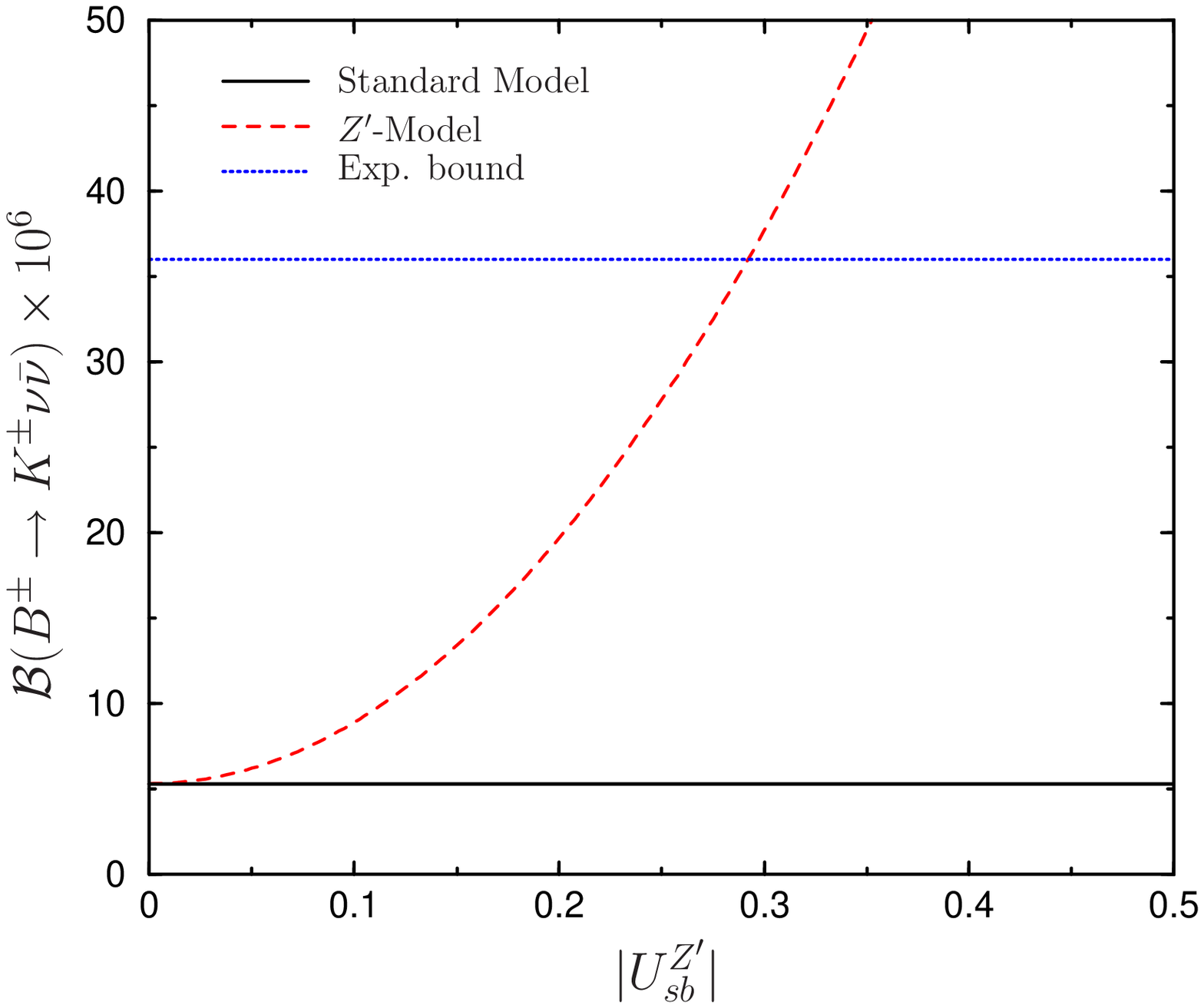,width=6cm}~~~&
~~~\psfig{file=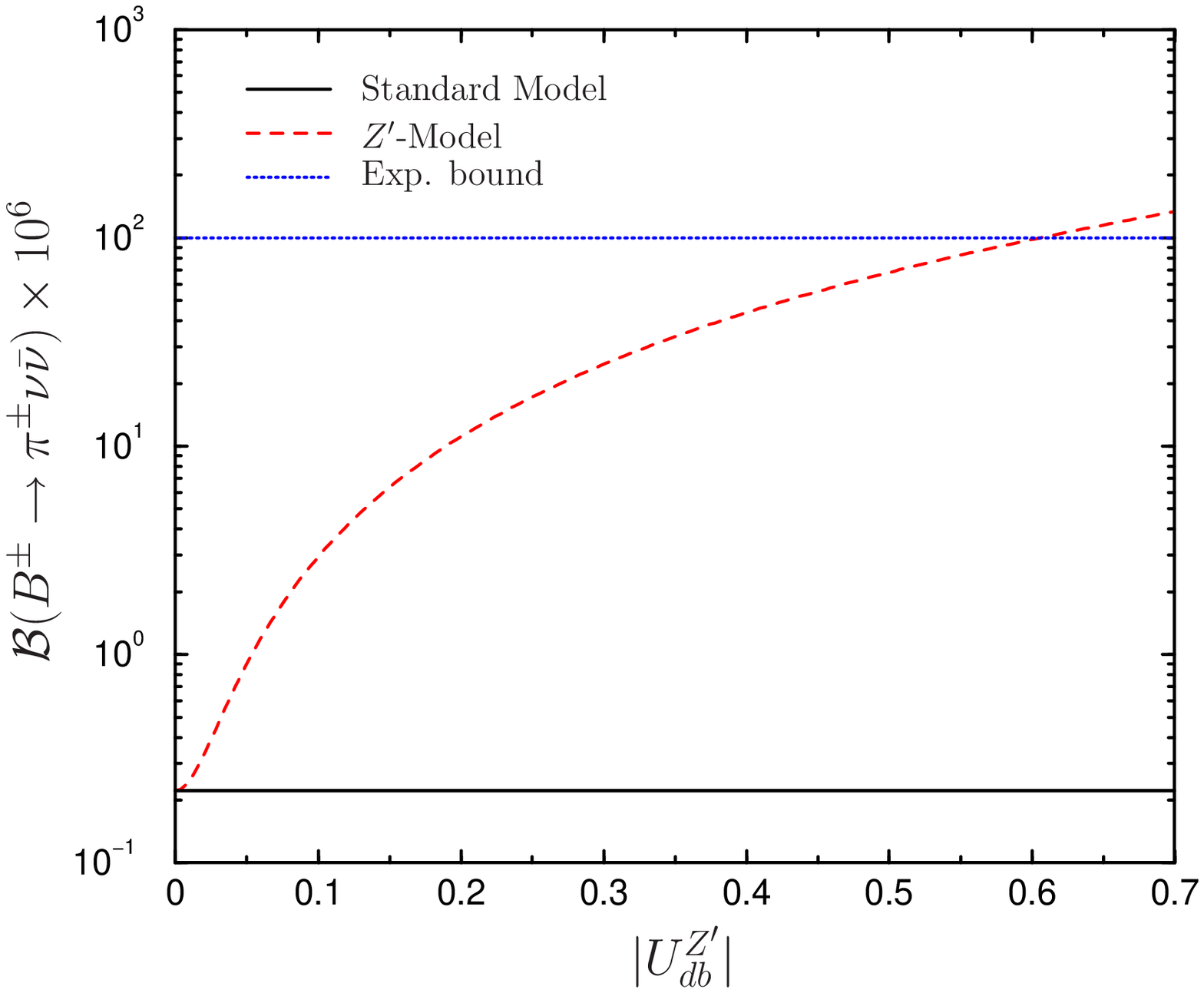,width=6cm}~~~\\[-0.0ex]
\textbf{(a)}&\textbf{(b)}
\end{tabular}
\vspace*{1pt}
\caption{ \label{fig1}
Branching ratios for (a)~$B^\pm \to K^\pm \nu\bar{\nu}$ and
(b)~$B^\pm \to \pi^\pm \nu\bar{\nu}$,
where $\nu$ can be the ordinary SM neutrinos or right-handed neutrinos.
}
\end{center}
\end{figure}
%%%%%%%%%%%%%%%%%%%%%%%%%%%%%%%%%%%%%%%%%%
%%%%%%%%%%%%%%%%%%%%%%%%%%%%%%%%%%%%%%%%%%

The leptophobic $Z^\prime$ model can yield same signals as
$B\to K\nu_{\rm SM} \bar{\nu}_{\rm SM}$ at detectors
via the production of a pair of right-handed neutrinos instead of
the ordinary SM neutrinos.
In Fig.~\ref{fig1}, we present our predictions for the BRs in the leptophobic
$Z^\prime$ model as a function of the effective coupling $|U_{qb}^{Z^\prime}|$,
where the mass of the $Z^\prime$ boson is assumed to be 700 GeV.
The solid and dotted lines represent the estimates in the SM and the current experimental bounds, respectively.
The dashed line denotes the expected BRs in the leptophobic $Z^\prime$ model.
In spite that we choose a specific mass for the $Z^\prime$ boson,
the present analysis can be easily translated through the corresponding changes
in the effective coupling $|U_{qb}^{Z^\prime}|$
for different $Z^\prime$ boson mass.
We extract the following constraints for the FCNC couplings from
Fig.~\ref{fig1}
\begin{equation}
|U_{sb}^{Z^\prime}| \leq 0.29 , ~~
|U_{db}^{Z^\prime}| \leq 0.61 ,
\label{Ubound}
\end{equation}
for $B\to K \nu\bar{\nu}$ and $B\to \pi \nu \bar{\nu}$ decays, respectively
\cite{Jeon:2006nq}.
The present exclusive mode gives more stringent bounds on the leptophobic
FCNC coupling compared with the inclusive $b\to s \nu\bar{\nu}$ decay
\cite{Leroux:2001fx}.

Recently, the Belle Collaboration has reported
upper limits on the production of the $K^\ast$ meson  with the missing
energy signal at the $B$ decay
where its BR is expected to be about 3 times larger than
that of the scalar meson production in the SM~\cite{:2006vg}.
It provides the constraint on the FCNC coupling
\begin{equation}
|U_{sb}^{Z^\prime}| \leq 0.66,
\end{equation}
which is larger than that in Eq.~(\ref{Ubound}).
At the super-$B$ factory, all four decay modes $B\to M\nu \bar{\nu}$
would be well measured and give more stringent bounds on the FCNC couplings.

The exclusive modes are much easier at the experimental detection than
the inclusive ones.
However, the exclusive modes have inevitable large theoretical
uncertainties from hadronic transition form factors.
In order to reduce hadronic uncertainties,
one can take ratios for  ${\cal B}(B\to M \nu \bar{\nu})$
to ${\cal B}(B\to M e \nu)$ for $M=\pi,\rho$ mesons \cite{Jeon:2006nq}.

\section{$B_s^0-\bar{B}_s^0$ Mixing
\label{sec4}}

The $Z^\prime$-exchanging $\Delta B = \Delta S = 2$
tree diagram contributes to the $B_s^0-\overline{B}_s^0$ mixing~\cite{Baek:2006bv}.
The mass difference $\Delta m_s$ of the mixing parameters then
read
\begin{eqnarray}
\Delta m_s = \Delta m_s^{\rm SM}
   \left|1 + R ~e^{2i \phi_{sb}^{Z^\prime}} \right|,
\end{eqnarray}
\begin{eqnarray}
R \equiv  \frac{2\sqrt{2} \pi^2}
     {G_F M_W^2 \left( V_{tb} V{_{ts}^\ast} \right)^2  S_0(x_t)}
     \frac{M_Z^2}{M_{Z^\prime}^2}
     \left|U_{sb}^{Z^\prime}\right|^2
     = 1.62 \times 10^3
       \left(\frac{700 ~\rm{GeV}}{M_{Z^\prime}}\right)^2
       \left|U_{sb}^{Z^\prime}\right|^2.
\end{eqnarray}

\begin{figure}
\begin{tabular}{cc}
~~~\psfig{file=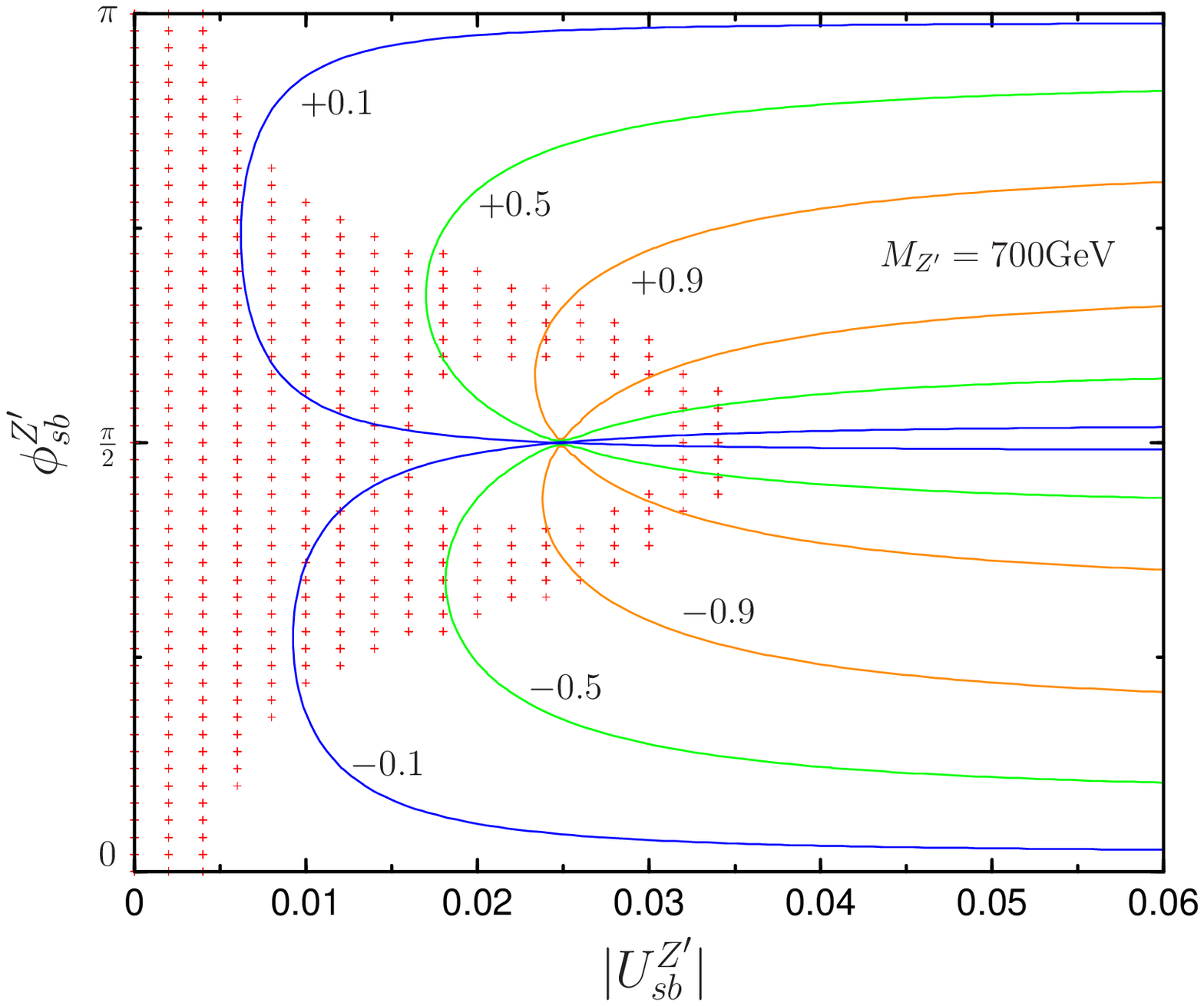,width=7cm}~~~&
~~~\psfig{file=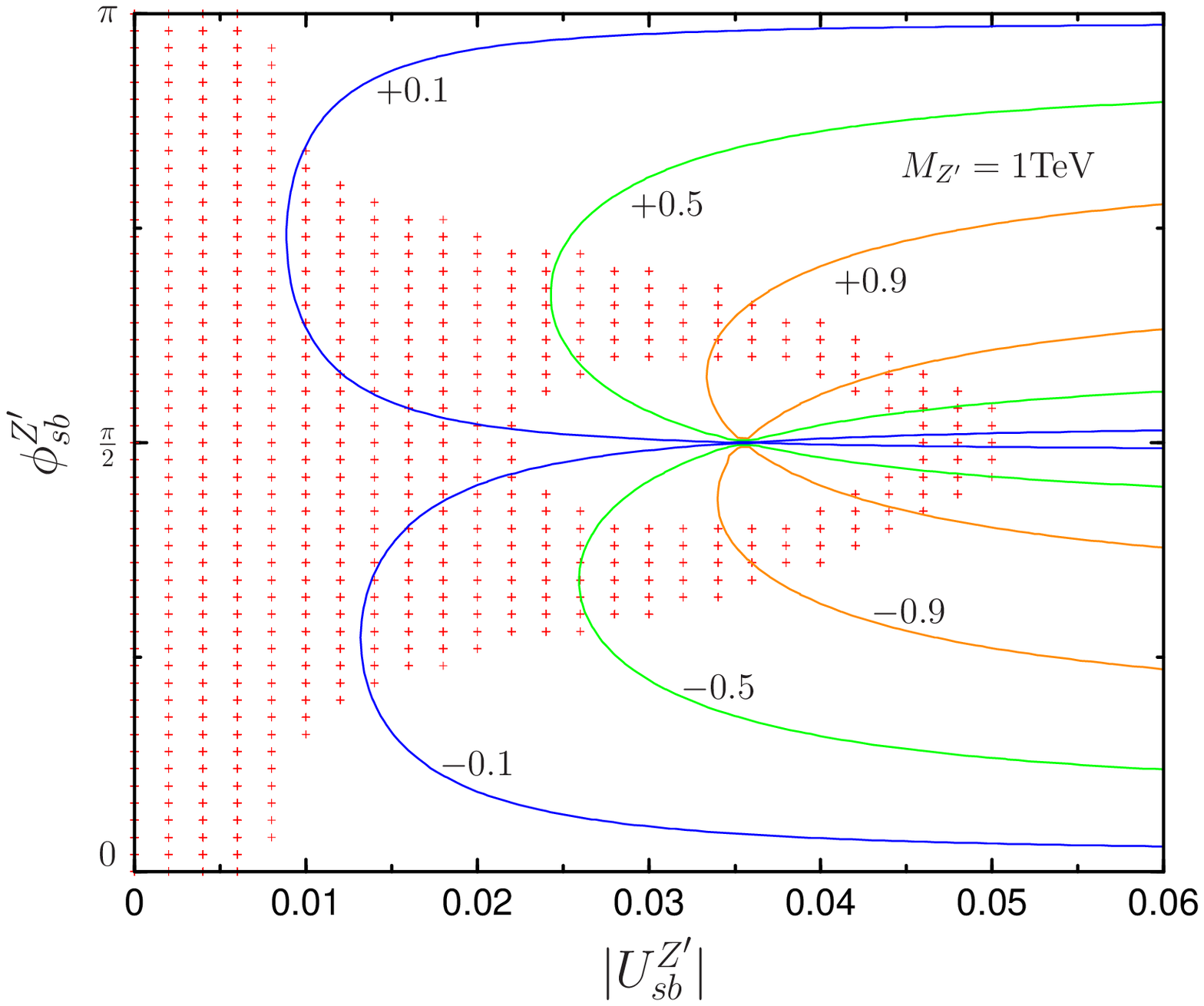,width=7cm}~~~\\[-0.5ex]
\textbf{(a)}&\textbf{(b)}
\end{tabular}
\vspace*{8pt} \caption{ \label{fig3} The allowed region in
($|U_{sb}^{Z^\prime}|$,$\phi_{sb}^{Z^\prime}$) plane for
(a)~$M_{Z^\prime}=700$ GeV and (b)~$M_{Z^\prime}=1$ TeV~. We used
(HP+JL)QCD result in \cite{lattice QCD} for the hadronic parameter.
Constant contour lines for the time dependent CP asymmetry
$S_{\psi\phi}$ in $B_s \to J/\psi~\phi$ are also shown. }
\end{figure}

In Figs.~\ref{fig3}, the allowed region in
($|U_{sb}^{Z^\prime}|$,$\phi_{sb}^{Z^\prime}$) plane is shown.
We obtain \begin{equation} |U_{sb}^{Z^\prime}|
\leq 0.0055 \qquad   \rm{for} ~M_{Z^\prime} = 700 ~ \rm{GeV},
\end{equation}
for $\phi_{sb}^{Z^\prime}=0$.
This bound is about two orders of magnitude stronger than
(\ref{Ubound}) obtained from exclusive semileptonic
$B \to M \nu \bar{\nu}$ decays.

The holes appear because they predict too small $\Delta m_s$.
For a given $M_{Z^\prime}$ we can see that large CP violating phase can enhance
the allowed coupling $|U_{sb}^{Z^\prime}|$ up to almost factor 10.
This shows the importance of the role played by CP violating phase
even in CP conserving observable such as  $\Delta m_s$.
As can be seen from Fig. 3(b),
irrespective of its phase $\phi_{sb}^{Z^\prime}$ value
\begin{equation}
|U_{sb}^{Z^\prime}| \leq 0.051 \qquad
                    \rm{for} ~M_{Z^\prime} = 1 ~ \rm{TeV}.
\end{equation}

The CP violating phase in $B_s^0 - \overline{B}_s^0$ mixing amplitude
can be measured at LHC in near future through the
time-dependent CP asymmetry in $B_s \to J/\psi~\phi$ decay
\begin{equation}
 \frac{ \Gamma \left(\overline{B}_s^0(t) \to J/\psi~ \phi \right)
       -\Gamma \left(B_s^0(t) \to J/\psi~ \phi \right)}
      { \Gamma \left(\overline{B}_s^0(t) \to J/\psi~ \phi \right)
       +\Gamma \left(B_s^0(t) \to J/\psi~ \phi \right)}
 \equiv S_{\psi\phi} \sin \left(\Delta m_s t\right).
\end{equation}
We note that although the final states are not CP-eigenstates, the time-dependent
analysis of the $B_s^0 \to J/\psi~ \phi$ angular distribution allows a clean
extraction of $S_{\psi\phi}$~\cite{angular}.
In the SM, $S_{\psi\phi}$ is predicted to be very small,
$S_{\psi\phi}^{\rm SM} =-\sin 2\beta_s =0.038 \pm 0.003$
$\left(\beta_s \equiv \arg \left[(V_{ts}^* V_{tb}) / (V_{cs}^*
V_{cb})\right]\right)$. If NP has an additional CP violating phase
$\phi_{sb}^{Z^\prime}$, however, the experimental value of
\begin{equation}
 S_{\psi\phi} = -\sin \left[ 2 \beta_s +
                            \arg \left(1 + R ~e^{2i \phi_{sb}^{Z^\prime}} \right)
                     \right]
\end{equation}
would be significantly different from the SM prediction.
Constant contour lines for $S_{\psi\phi}$ are also shown in Figs.~\ref{fig3}.
We can see that even with the strong constraint from the present $\Delta m_s$ observation,
large $S_{\psi\phi}$ are still allowed.

\section{Concluding Remarks
\label{sec5}}

In this talk, we have considered the leptophobic $Z^\prime$ model
with FCNC couplings.
Since the direct probe of the leptophobic $Z^\prime$ model is very difficult,
the exclusive $B\to M \nu \bar{\nu}$ decay are very adequate
to measure the FCNC coming form this model.
We have also showed that the recently measured mass difference
$\Delta m_s$ of $B_s^0-\overline{B}_s^0$ system
can constrain this kind of models very efficiently.
Although the bounds on the coupling estimated from the latter
are about two orders of magnitudes stronger than those from the former,
both measurements are complementary.

%%%%% acknowledgement %%%%%
\vspace*{12pt}
\noindent
{\bf Acknowledgement}

\noindent C.S.K. is supported by the KRF Grant funded by the Korean Government (MOEHRD) No. KRF-2005-070-C00030.

\vspace*{6pt}

\noindent

%%%%% references %%%%%


\begin{thebibliography}{99}

%\cite{Kim:2005jp}
\bibitem{Kim:2005jp}
  S.~Baek, P.~Hamel, D.~London, A.~Datta, and D.~A.~Suprun,
  %``The B --> pi K puzzle and new physics,''
  Phys.\ Rev.\ D {\bf 71}, 057502 (2005)
  [arXiv:hep-ph/0412086];
  %%CITATION = HEP-PH 0412086;%%
  C.~S.~Kim, S.~Oh, and C.~Yu,
  %``A critical study of the B --> K pi puzzle,''
  Phys.\ Rev.\ D {\bf 72}, 074005 (2005)
  [arXiv:hep-ph/0505060];
  %%CITATION = HEP-PH 0505060;%%
  S.~Baek,
  %``New physics in B --> pi pi and B --> pi K decays,''
  JHEP {\bf 0607}, 025 (2006)
  [arXiv:hep-ph/0605094].
  %%CITATION = HEP-PH 0605094;%%


%\cite{Wu:2006ur}
\bibitem{Wu:2006ur}
  Y.~L.~Wu, Y.~F.~Zhou, and C.~Zhuang,
  %``Implications of new data in charmless B decays,''
  arXiv:hep-ph/0609006 and references therein.
  %%CITATION = HEP-PH 0609006;%%

\bibitem{D0}
  V.~M.~Abazov {\it et al.}  [D0 Collaboration],
  %``First direct two-sided bound on the B/s0 oscillation frequency,''
  Phys.\ Rev.\ Lett.\  {\bf 97}, 021802 (2006)
  [arXiv:hep-ex/0603029].
  %%CITATION = HEP-EX 0603029;%%

%\cite{Gomez-Ceballos:2006qm}
\bibitem{Gomez-Ceballos:2006qm}
   A.~Abulencia {\it et al.} [CDF - Run II Collaboration],
  %``Measurement of the B/s0 anti-B/s0 oscillation frequency,''
  Phys.\ Rev.\ Lett.\  {\bf 97}, 062003 (2006)
  [arXiv:hep-ex/0606027].
  %%CITATION = HEP-EX 0606027;%%
%  G.~Gomez-Ceballos and J.~Piedra  [CDF Collaboration],
%  %``B mixing and lifetimes at the Tevatron,''
%  eConf {\bf C060409}, 011 (2006)
%  [arXiv:hep-ex/0606048].
%  %%CITATION = HEP-EX 0606048;%%

\bibitem{MSSM}
   S.~Baek,
 %  ``B/s - anti-B/s mixing in the MSSM scenario with large flavor mixing in  the
  %LL/RR sector,''
  JHEP {\bf 0609}, 077 (2006)
  [arXiv:hep-ph/0605182];
  %%CITATION = HEP-PH 0605182;%%
  R.~Arnowitt, B.~Dutta, B.~Hu, and S.~Oh,
%   ``B/s - anti-B/s mixing and its implication for b --> s transitions in
%  supersymmetry,''
  Phys.\ Lett.\ B {\bf 641}, 305 (2006)
  [arXiv:hep-ph/0606130].
  %%CITATION = HEP-PH 0606130;%%


\bibitem{RS-kim}
S.~Chang, C.~S.~Kim, and J.~Song,
  %``Constraint of B/d,s0 anti-B/d,s0 mixing on warped extra-dimension model,''
  arXiv:hep-ph/0607313.
  %%CITATION = HEP-PH 0607313;%%


\bibitem{Lopez:1996ta}
  J.~L.~Lopez and D.~V.~Nanopoulos,
  %``Leptophobic $Z'$ in stringy flipped SU(5),''
  Phys.\ Rev.\ D {\bf 55}, 397 (1997)
  [arXiv:hep-ph/9605359].
  %%CITATION = HEP-PH 9605359;%%

\bibitem{Rizzo:1998ut}
T.~G.~Rizzo,
  %``Gauge kinetic mixing and leptophobic Z' in E(6) and SO(10),''
  Phys.\ Rev.\ D {\bf 59}, 015020 (1999)
  [arXiv:hep-ph/9806397].
  %%CITATION = HEP-PH 9806397;%%

\bibitem{Leroux:2001fx}
  K.~Leroux and D.~London,
  %``Flavour-changing neutral currents and leptophobic Z' gauge bosons,''
  Phys.\ Lett.\ B {\bf 526}, 97 (2002)
  [arXiv:hep-ph/0111246].
  %%CITATION = HEP-PH 0111246;%%

\bibitem{Jeon:2006nq}
  J.~H.~Jeon, C.~S.~Kim, J.~Lee, and C.~Yu,
  %``Exclusive B $\to$ M nu anti-nu (M = pi, K, rho, K*) decays and leptophobic
  %Z' model,''
  Phys.\ Lett.\ B {\bf 636}, 270 (2006)
  [arXiv:hep-ph/0602156].
  %%CITATION = HEP-PH 0602156;%%

%\cite{:2006vg}
\bibitem{:2006vg}
   K.~Abe {\it et al.} [Belle Collaboration],
  %``Search for B0 --> K*0 nu anti-nu using one fully reconstructed B meson,''
  arXiv:hep-ex/0608047.
  %%CITATION = HEP-EX 0608047;%%

%\cite{Abe:2005bq}
\bibitem{Abe:2005bq}
  K.~Abe {\it et al.}  [Belle Collaboration],
  %``Search for B $\to$ tau nu and B $\to$ K nu anti-nu decays with a fully
  %reconstructed B at belle,''
  arXiv:hep-ex/0507034.
  %%CITATION = HEP-EX 0507034;%%

%\cite{Aubert:2004ws}
\bibitem{Aubert:2004ws}
  B.~Aubert {\it et al.}  [BABAR Collaboration],
  %``A search for the decay B+ $\to$ K+ nu anti-nu,''
  Phys.\ Rev.\ Lett.\  {\bf 94}, 101801 (2005)
  [arXiv:hep-ex/0411061].
  %%CITATION = HEP-EX 0411061;%%

\bibitem{Baek:2006bv}
  S.~Baek, J.~H.~Jeon, and C.~S.~Kim,
  %``B/s0 - anti-B/s0 mixing in leptophobic Z' model,''
  Phys.\ Lett.\ B {\bf 641}, 183 (2006)
  [arXiv:hep-ph/0607113];
  %%CITATION = HEP-PH 0607113;%%

\bibitem{lattice QCD}
%\bibitem{Aoki:2003xb}
  S.~Aoki {\it et al.}  [JLQCD Collaboration],
  %``B0 anti-B0 mixing in unquenched lattice QCD,''
  Phys.\ Rev.\ Lett.\  {\bf 91}, 212001 (2003)
  [arXiv:hep-ph/0307039];
  %%CITATION = HEP-PH 0307039;%%
%\bibitem{Gray:2005ad}
  A.~Gray {\it et al.}  [HPQCD Collaboration],
  %``The B meson decay constant from unquenched lattice QCD,''
  Phys.\ Rev.\ Lett.\  {\bf 95}, 212001 (2005)
  [arXiv:hep-lat/0507015];
  %%CITATION = HEP-LAT 0507015;%%
%\bibitem{Okamoto:2005zg}
  M.~Okamoto,
  %``Full determination of the CKM matrix using recent results from lattice
  %QCD,''
  PoS {\bf LAT2005}, 013 (2006)
  [arXiv:hep-lat/0510113];
  %%CITATION = HEP-LAT 0510113;%%
%\bibitem{Ball:2006xx}
  P.~Ball and R.~Fleischer,
  %``Probing new physics through B mixing: Status, benchmarks and prospects,''
  arXiv:hep-ph/0604249.
  %%CITATION = HEP-PH 0604249;%%


\bibitem{angular}
  A.~S.~Dighe, I.~Dunietz, and R.~Fleischer,
   %``Extracting CKM phases and B/s anti-B/s mixing parameters from  angular
  %distributions of non-leptonic B decays,''
  Eur.\ Phys.\ J.\ C {\bf 6}, 647 (1999)
  [arXiv:hep-ph/9804253];
  %%CITATION = HEP-PH 9804253;%%
  I.~Dunietz, R.~Fleischer, and U.~Nierste,
  %``In pursuit of new physics with B/s decays,''
  Phys.\ Rev.\ D {\bf 63}, 114015 (2001)
  [arXiv:hep-ph/0012219].
  %%CITATION = HEP-PH 0012219;%%

\end{thebibliography}
\end{document}